# Objektorientierte Graphendarstellung von Simulink-Modellen zur einfachen Analyse und Transformation

Von Carsten Kolassa, David Dieckow, Michael Hirsch, Uwe Creutzburg, Christian Siemers, Bernhard Rumpe [1]

## 1 Einleitung

Für die Erstellung von cyberphysikalischen Systemen ist die modellbasierte Entwicklung ein Standardwerkzeug, da es die Erstellung komplexer Systeme vereinfacht und übersichtlich ermöglicht. Durch die Simulation lassen sich Verifikationen bereits in sehr frühen Entwicklungsstadien durchführen.

Eines der am häufigsten eingesetzten Werkzeuge in diesem Bereich ist MATLAB/Simulink. In Simulink können die ausführbaren Modelle durch Blockschaltbilder und Zustandsdiagramme erstellt werden. Diese ausführbaren Modelle können in vielen verschiedenen Anwendungsbereichen eingesetzt werden. Modelle können z. B. automatisch in Quellcode überführt werden, um sie auf einem Steuergerät zu verwenden, oder es kann eine Testeinrichtung mit simulierten Umgebungswerten stimuliert werden.

Für viele Anwendungsbereiche ist es wünschenswert Analysen auf dem und Transformationen auf das Simulink-Modell durchzuführen. Analysen können z. B. das Erkennen von speziellen Mustern[2] oder Klonen[3] sein. Unter der Transformation versteht man z.B. das Substituieren eines bestimmten Teils des Modells durch ein anderes Modell.

Die von Matlab bereitgestellte API (Application Programming Interface) ist jedoch ungeeignet, um derartige Operationen auf ein Modell aus MATLAB heraus durchführen zu können.

Im Rahmen dieser Arbeit wird das **MAnTrAS**-System (**M**ATLAB **An**alyse und **Tr**ansfomations **A**PI für **S**imulink) vorgestellt, mit dem die angesprochenen Analysen und Transformationen komfortabel möglich sind. Um die Komfortabilität zu gewährleisten, wird das Modell in eine objektorientierte Graphenrepräsentation überführt. Mittels dieser Beschreibungsform lassen sich graphentheoretische Algorithmen einfacher anwenden als direkt auf das Simulink-Modell. Das System ermöglicht sowohl Transformationen eines Simulink-Modells in eine Graphenrepräsentation als auch die Rücktransformationen.

Dieses Papier ist wie folgt gegliedert: In Abschnitt 2 wird der aktuelle Stand der Technik dargelegt und aufgezeigt, welche Ansätze bisher verfolgt wurden. Auf der Basis dieser Grundlagen wird in Abschnitt 3 das neue Domänenmodell vorgestellt. Darauf aufbauend wird in Abschnitt 4 auf die Implementierung von MAnTrAS und dessen Eigenschaften eingegangen. Im folgenden Abschnitt 5 werden Anwendungen beschrieben, welche die Vorteile des Systems im Einsatz darstellen. Abschließend wird in Abschnitt 6 ein Fazit gezogen.

---

[1] M.Sc. Carsten Kolassa, Lehrstuhl für Software Engineering RWTH Aachen; M.Sc. David Dieckow, Institute for Applied Computer Science, Fachhochschule Stralsund; M.Sc. Michael Hirsch, Zentrale Elektronik-Entwicklung (HE-2) Volkswagen Aktiengesellschaft, Braunschweig; Prof. Dr. Uwe Creutzburg Fachbereich Elektrotechnik und Informatik, Fachhochschule Stralsund; Prof. Dr. Christian Siemers, Institut für Informatik, TU-Clausthal; Prof. Dr. Bernhard Rumpe, Lehrstuhl für Software Engineering RWTH Aachen;

[2] Muster sind bestimmte Blockkonstellationen innerhalb des Modells

[3] Klone sind einzelne Blöcke oder Gruppen von Blöcke, welche mehrfach auftreten

## 2 Stand der Technik

Im folgenden Abschnitt grenzen wir MAnTrAS von bestehenden Arbeiten ab und stellen die bestehende MATLAB API dar, zu welcher MAnTrAS eine Verbesserung darstellt.

### 2.1 Bestehende Arbeiten

Arbeiten wie [2], [3] und [5] verfolgen ein ähnliches Ziel wie MAnTrAS: das Analysieren und Transformieren von Simulink-Modellen. So beschreibt [5] sehr umfangreich Metriken, die zur Bewertung von Simulink-Modellen bezüglich ihrer Qualität verwendet werden können, sowie Ansätze für deren Implementierung. Mit INProVE [3] wird ein weiteres Tool zur Qualitätsmessung vorgestellt. Es setzt dabei auf ausgereifte Mustererkennung um eine Qualitätsanalyse des Modells durchzuführen. Sowohl [5] als auch [3] beschränken sich dabei auf die reine Analyse von MATLAB Modellen, Transformationen sind nicht möglich.
Das MATE Projekt [2] geht einen Schritt weiter und bietet neben der Erkennung von Mustern die Möglichkeit Modelle zu transformieren. Dabei werden vom Benutzer Transformationsregeln erstellt, die Substitutionen auf Blöcke oder Muster von Blöcken anwenden. Um Vergleiche zwischen Simulink-Modellen durchzuführen, wurde MATE um den SiDiff-Algorithmus [9] erweitert, der auf die Analyse von UML-Modellen basiert. MATE hat dabei den Zweck Verletzungen von Modellierungsrichtlinien aufzudecken und wenn möglich zu beheben. Es unterscheidet sich dabei von MAnTrAS, welches allgemein der Analyse und Transformation dient.
In [4] wird ein Prototyp zur Erkennung und Korrektur von Verstößen gegen vorgegebene Richtlinien für Simulink-Modelle vorgestellt. Der Prototyp setzt dabei auf die FUJABA Tool Suite [8] auf, die über ein M-Script Interface in der Lage ist, Matlab-Code auszuführen. Dabei werden vorher definierte Muster erkannt und transformiert.
MAnTrAS unterscheidet sich von den bestehenden Arbeiten insoweit, dass es nicht vorgegeben Transformationen oder Analysen anbietet, sondern weiterhin dem Benutzer ein Werkzeug in die Hand gibt, um eigene komplexe Analysen und Transformationen zu erstellen.

### 2.2 Bestehende MATLAB-Simulink API

Simulink ist ein in MATLAB integriertes Werkzeug zur hierarchischen Modellierung von Systemen mittels graphischer Elemente. MATLAB stellt eine API bereit um auf Simulink Modelle zuzugreifen. Die bestehende MATLAB API ist jedoch prozedural konzipiert, wodurch nicht graphenbasiert programmatisch auf ein Simulink-Modell zugegriffen werden kann. Um beispielsweise die in einem System enthaltenen Blöcke zu erhalten, wird die Funktion "find_system" verwendet, die eine Liste von Blöcken zurückliefert. Möchte man nun Informationen über die Verbindungen der Blöcke untereinander erhalten, muss eine weitere Funktion verwendet werden. Mithilfe des Kommandos "get_param(BlockName, 'LineHandles')" erhält man eine Referenz auf die Linien, die mit einem Block verbunden sind. Um weitere Informationen, wie Quell- und Zielport zu erhalten, muss wiederum die "get_param"-Funktion verwendet werden. Das zeigt, dass über Matlab nur sehr umständlich auf den Simulink-Graphen zugegriffen werden kann. Abläufe, wie das einfache Ermitteln von Vorgänger und Nachfolger eines Blocks sind bereits sehr umständlich zu implementieren. Die Erstellung von komplexen Analyse- und Transformationsalgorithmen wird damit aufwändig und fehleranfällig.

# 3 Graph basiertes Domänenmodell

Wie im vorherigen Kapitel beschrieben ist die momentane API nicht graphenbasiert. Das semantische Model von Simulink Diagrammen ist aber ein gerichteter Graph mit eigenständigen Mehrfachkanten, wobei die Blöcke (Block) Knoten darstellen, welche über Linien (Line), die Kanten abbilden, miteinander verbunden sind (siehe Abbildung 1).

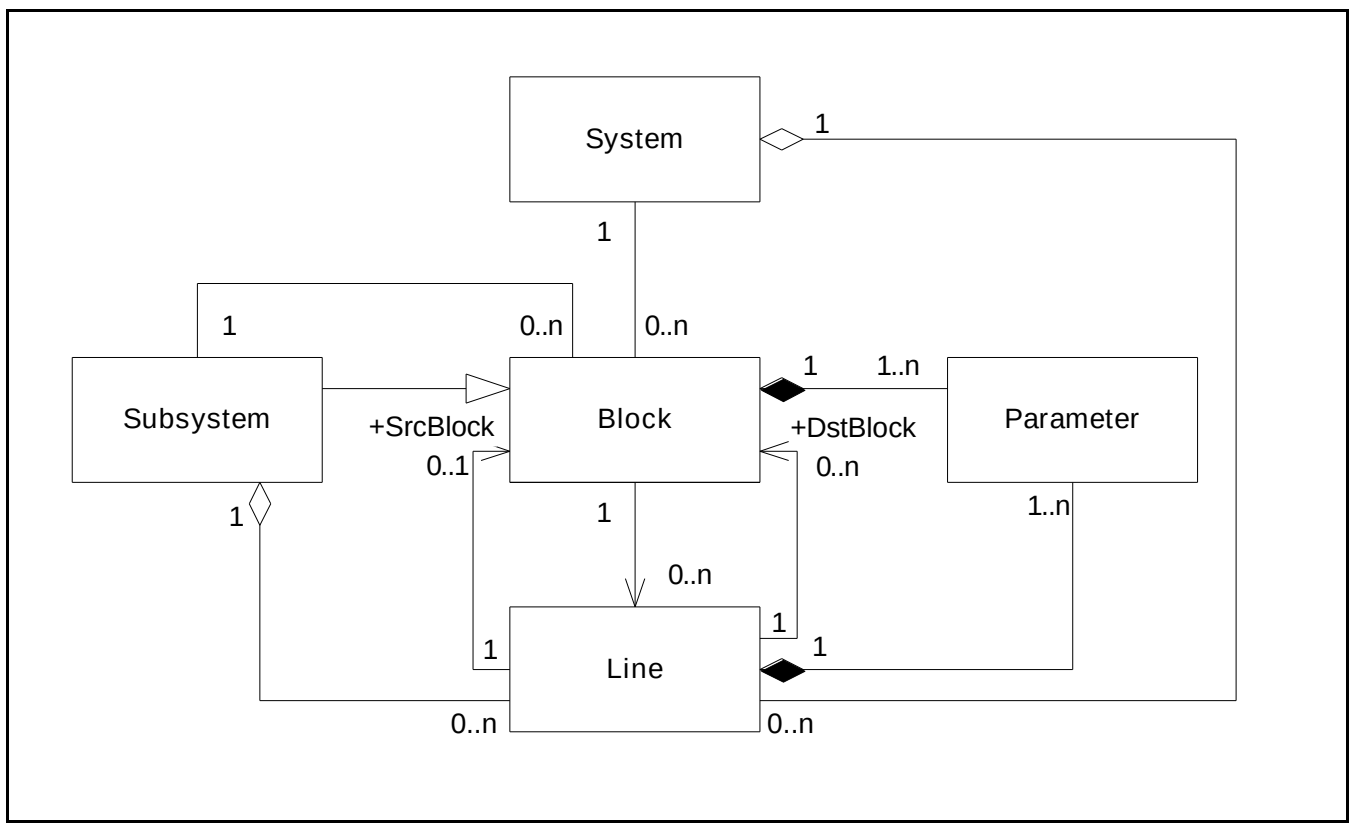

Abbildung 1 Simulink Domänenmodell in UML

Hierarchien werden in Simulink durch Subsysteme dargestellt, welche ihrerseits einen abgekapselten Teilgraphen bilden und weitere Blöcke sowie Linien enthalten können. Dabei unterscheidet man virtuelle und nicht virtuelle Subsysteme. Letztere werden als Einheit ausgeführt (atomic execution), die ersteren dienen nur der hierarchischen Abgrenzung von Blöcken [10].
Blöcke sind mit Eigenschaften (Parameter) versehen. Sie bestimmen beispielsweise die Farbe sowie die funktionale Ausprägung eines Blocks bzw. einer Linie. Die Parameter werden deshalb als Eigenschaft eines Knotens oder einer Kante betrachtet und nicht gesondert im Graphen abgebildet.

# 4 Implementierung

Im Folgenden gehen wir auf die Implementierungsdetails ein und beschreiben den Aufbau der Architektur von MAnTrAS.

## 4.1 Überblick

Für die Implementierung haben wir ein objektorientiertes Design gewählt. Dieser Designansatz allein ermöglicht es uns eine effizientere und intuitivere API anzubieten. Die Implementierung orientiert sich an dem vorgestellten Domänenmodell. Nachfolgendes Klassendiagram (Abbildung 2) zeigt die implementierten Klassen. Jede Klasse im Domänenmodell hat eine korrespondierende Klasse in der Implementierung.
Der *GraphCreator* greift mit Hilfe der MATLAB zu Simulink API (Abschnitt 2.2) auf das Simulink Modell zu und erzeugt aus ihm einen Graphen, dessen Knoten Blöcke darstellen, welche durch Kanten verknüpft sind, die durch die *Connections* abgebildet werden.

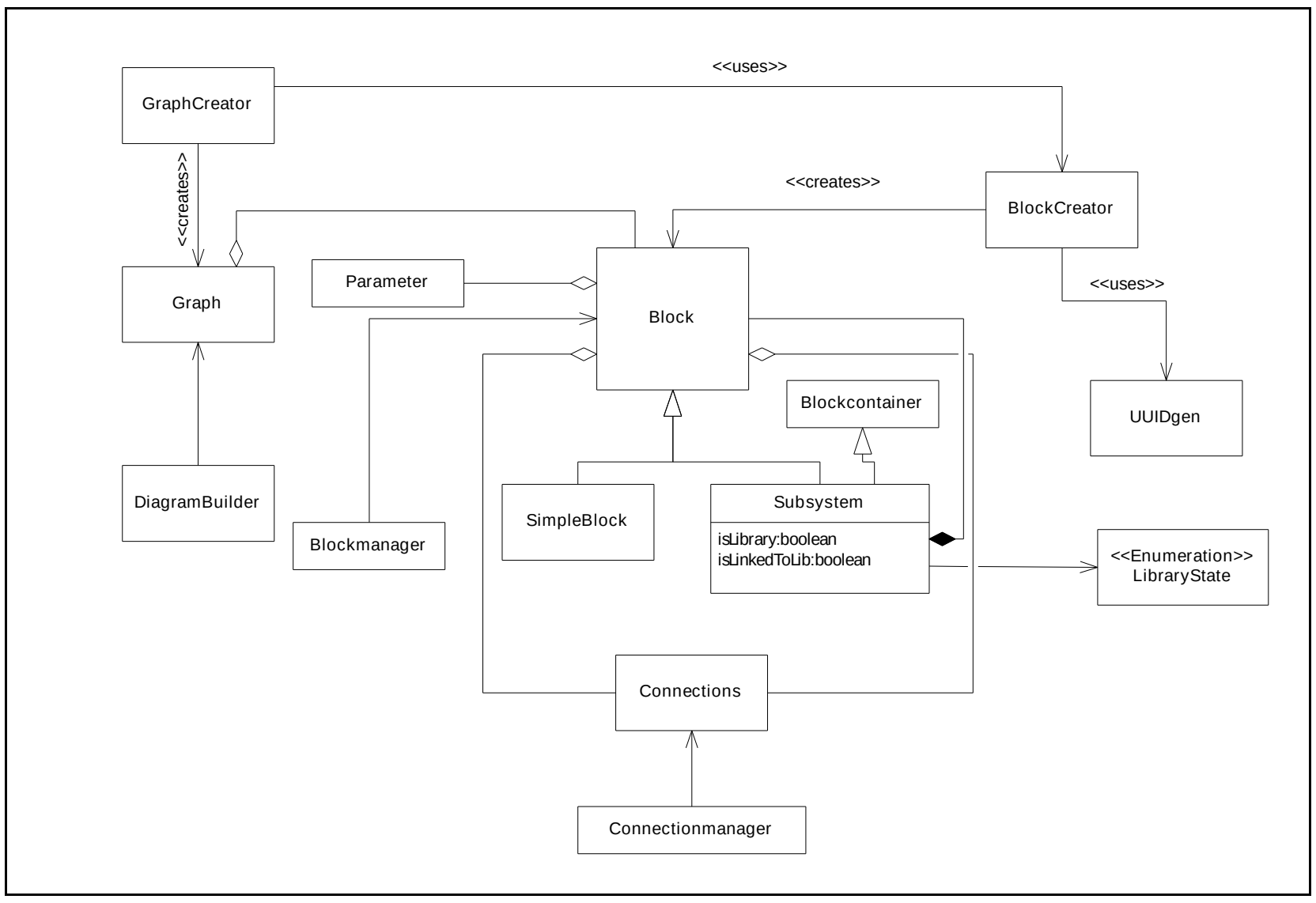

Abbildung 2 UML-Klassendiagram der Architektur von MAnTrAS

Blöcke spezialisieren sich in *SimpleBlock* und *Subsystem* wobei *SimpleBlock* einen atomaren Block repräsentiert. Objekte der Klasse *Subsystem* können wiederum selbst Blöcke enthalten, auf welche sie verweisen. Diese Eigenschaft erben sie von ihrer Oberklasse *BlockContainer*.
Bei der Verarbeitung von Subsystemen wird nicht berücksichtigt ob sie virtuell oder nicht virtuell sind, da bei der Transformation und Analyse die Eigenschaft der atomic execution keine Rolle spielt.
In der Architektur von MAnTrAS wird zwischen eingehenden und ausgehenden Verbindungen unterschieden, wodurch die Richtung des Graphen abgebildet wird.
Sowohl Blöcke als auch Verbindungen werden jeweils durch einen eigenen Manager verwaltet, welcher als Singleton implementiert ist. Der Manager hält dabei eine Referenz auf die verwalteten Entitäten. Des Weiteren implementiert er Suchfunktionen, über die sich Blöcke und Verbindungen nach ihren Eigenschaften finden lassen.
Entitäten, welche durch Manager verwaltet werden, erhalten eine generierte, 128 Bit breite, UUID[4]. Dies ist notwendig, da die intern in MATLAB verwendeten IDs für Blöcke und Verbindungen nach einer Transformation und der dadurch bedingten Neuerzeugung von Blöcken oder Verbindungen neu vergeben werden. Die UUIDs gewährleisten eine eindeutige Identifizierung von Blöcken unabhängig von der von Simulink vergebenen ID. UUIDs bleiben auch über die Modelltransformationen hinweg erhalten. Im Gegensatz dazu werden die Simulink Block-Referenzen (Block-Handle) durch Transformation und Rücktransformation verändert.
Eine weitere wichtige Funktion von MAnTrAS ist die Rücktransformation eines Graphen in ein Simulink-Modell. Sie wird durch den *DiagramBuilder* durchgeführt, wobei die in den Parametern gespeicherten Eigenschaften erhalten bleiben. Wird nur eine Transformation in ein MAnTrAS Modell und zurück durchgeführt, ohne eine Veränderung am Modell durchzuführen, bleibt die semantische Äquivalenz gewährleistet, da alle Verbindungen und Parameter identisch zum Ausgangsmodel sind. Ausgangsmodell und generiertes Simulink-Modell unterscheiden sich nur in der graphischen Anordnung der Blöcke, sowie in den neu erzeugten Block-Referenzen.

[4] Universal Unique Identifier

## 4.2 Visitor in objektorientiertem M-Skript

Ein zentraler Bestandteil von MAnTrAS sind die „Visitor-Pattern“ [7]. Sie sind ein bekanntes und oft genutztes Entwurfsmuster in der objektorientierten Softwareentwicklung und bieten sich zur Verarbeitung von Graphen an. Bisherige Implementierungen von „Visitor-Pattern“ in MATLAB [1] sind nicht objektorientiert, da dies bis Matlab Release 2008a [6] nicht möglich war. Für MAnTrAS haben wir die Implementierung eines objektorientierten „Visitor-Pattern“ in M-Skript entworfen.

Normalerweise werden bei der Implementierung des „Visitor-Pattern‘s“ Methoden überladen. Da MATLAB das Überladen von Methoden nicht unterstützt mussten wir ein leicht abgewandeltes Konzept verwenden. Dabei wird zur Laufzeit der Typ des Objektes, welches gerade besucht wird, überprüft und in Abhängigkeit davon die entsprechende Methode aufgerufen. Da alle Typen von Klassen, welche in unserem Klassenmodell verwendet werden, bekannt sind, haben wir diesen Mechanismus in eine Oberklasse ausgelagert:

```
classdef Visitor < AbstractVisitor
    methods
        function run(obj,graph)
                    //Runner as in normal visitors
        end
        function result = visit(obj,element,graph)
            if (isa(element,'Subsystem'))
                 result = obj.visitSubsystem(element,graph);
            elseif (isa(element,'SimpleBlock'))
                result = obj.visitBlock(element,graph);
            end
        end
        function result = visitSubsystem(obj,element,graph)
            result=0; //Will be filed in the child classes
        end
        function result = visitBlock(obj,element,graph)
            result=0; //Will be filed in the child classes
        end
    end
end
```

Listing 1: „Visitor-Pattern“ in Matlab Code umgesetzt

Listing 1 zeigt die Oberklasse, von welcher sich alle konkreten Implementierungen von „Visitor-Pattern“ ableiten. Bei der bestehenden Implementierung in anderen Systemen, beispielsweise MATLAB BGL [1], wird ein „Visitor-Pattern“ nicht in einem Objekt gekapselt sondern durch zwei Funktionen repräsentiert. Dies macht es schwieriger das „Visitor-Pattern“ zu verwenden da immer beide Funktionen selbst implementiert werden müssen.

Für MAnTrAS ist es aber von zentraler Wichtigkeit, dass das „Visitor-Pattern“ leicht umzusetzen ist, da Anwender von MAnTrAS eigene Erweiterungen mit seiner Hilfe implementieren. Solche Erweiterungen können Modeltransformationen, aber auch Analysen des Modells sein.

# 5 Anwendungen

Neben der reinen MAnTrAS Implementierung wurden zahlreiche Anwendungen in Form von Analysen und Transformationen entwickelt. In diesem Abschnitt werden beispielhaft einzelne Anwendungen vorgestellt.

## 5.1 Analysen

Analysen von Graphen stellen sich prinzipiell als Mustererkennung dar. Bei einer Mustererkennung werden spezielle Blockkonstellationen bzw. spezielle Eigenschaften von diesem

Konstellationen gesucht und analysiert. Bereits in [5] werden zahlreiche Analysen für Simulink-Modelle vorgestellt, von denen in dieser Arbeit einige aufgegriffen und implementiert wurden. In diesem Kapitel beispielhaft auf die Implementierung verschiedener Analysen eingegangen.

### 5.1.1 Loop Detection

Die *Loop Detection* hat zur Aufgabe Zyklen in einem Graphen zu finden. Zyklen sind dabei Pfade in einem Graphen, welche auf ihren Ursprung zurückführen. Für weitere Analysen können Zyklen dazu führen, dass diese nicht durchführbar sind. Ein Beispiel hierfür ist die Suche nach parallelen Pfaden. Daher ist es wichtig, im Vorfeld Zyklen zu erkennen und mittels einer geeigneten Transformation zu eliminieren. Für die *Loop Detection* müssen von einem Startpunkt die Nachfolger durchgegangen werden und die durchlaufenden Pfade abgespeichert werden. Wird beim Durchlaufen der Nachfolger ein Block zweimal besucht, wird dieser Zyklus erkannt und kann später weiterverarbeitet werden.

### 5.1.2 Parallel Path

Parallele Pfade spielen in der Analyse eine große Rolle. In der modellbasierten Entwicklung stehen parallele Pfade für mindestens zwei Verarbeitungswege, welche einen gemeinsamen Start- und Endpunkt besitzen. Die Problematik bei diesen Pfaden besteht darin, dass sie, in speziellen Anwendungsgebieten, eine Echtzeitfähigkeit verlangen und daher zeitlich voneinander abhängig sind. Dabei muss sichergestellt werden, dass bei einer Zusammenführung von parallelen Pfaden die Werte zum richtigen Zeitpunkt eintreffen (Synchronität). Daher ist es bei der Veränderung des Modells wichtig frühzeitig solche Abhängigkeiten zu erkennen.

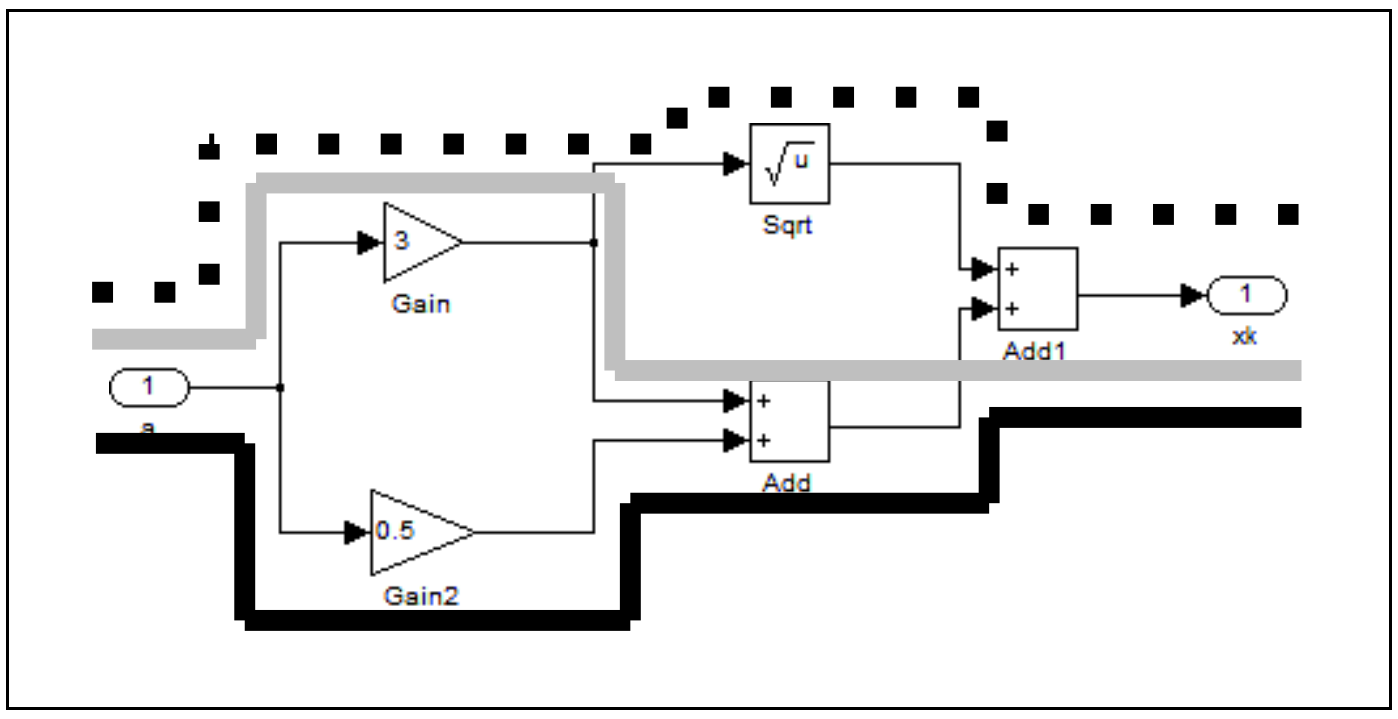

Abbildung 3: Ein Modul mit insgesamt sieben Pfaden von denen hier drei parallele gekennzeichnet (schwarz, grau, schwarz gestrichelt) sind.

Zur Erkennung paralleler Pfade wird zunächst die Menge aller Pfade des Systems ermittelt. Ein Pfad beginnt dabei immer an einem Block der mehrere Ausgänge besitzt und endet an einem Block mit mehreren Eingängen. Aus dieser Menge werden anschließend die Pfade extrahiert, die sowohl den gleichen Start-, als auch den gleichen End-Block besitzen, damit erhält man die Menge aller parallelen Pfade. Dabei ist zu beachten, dass vor der Pfadanalyse alle Zyklen im Modell durch geeignete Transformationen aufgelöst werden müssen.

### 5.1.3 Blocks on Path

Für einige Anwendungen kann es sinnvoll sein, die Anzahl eines bestimmten Blocktyps innerhalb eines Pfades - genau genommen sogar mehrere Pfade, also von einem Block zu einem anderen - zu analysieren. In der modelbasierten Hardwareentwicklung ist beispielsweise die Anzahl der Delays ausschlaggebend für die Antwortzeit eines Moduls (Algorithmus). Gleichzeitig kann überprüft werden, ob in einem Modul, auf allen Pfaden die gleiche Anzahl an Delays vorhanden sind, dies stellt die Synchronität der Ausgangswerte sicher.
Diese Analyse arbeitet ähnlich wie die Schleifendetektion (siehe Abschnitt 5.1.1) indem systematisch alle Pfade von einem Start zu einem Zielblock abgelaufen werden, auf dem Weg vom Start zum Ziel werden dabei die gewünschten Blöcke gezählt.

### 5.1.4 Clone Detection

Eine weitere Anwendung von MAnTrAS besteht in der Erkennung von Klonen. Dazu kann derselbe Ansatz wie in [11] verwendet werden. Das System wird zunächst normalisiert d.h. es werden die Subsysteme aufgelöst, sodass ein Modell mit flacher Hierarchie entsteht. Anschließend werden alle möglichen Konstellationen von Blöcken und deren Verbindungen gebildet und mithilfe einer breath-first-search (BFS) durchsucht, um mögliche Klone zu finden.
Unsere Lösung unterscheidet sich dabei von der in [11] beschriebenen, da MAnTrAS es erlaubt die komplette Implementierung in MATLAB durchzuführen, während die Implementierung in [11] ein externes Programm ist, welches auf Simulink Modelle angewendet wird.

### 5.1.5 Benutzerdefinierte Analysen durch Zusatzinformationen

Über die Analysen von graphentheoretischen Problemen hinaus, ist es möglich eigene Analysen durchzuführen. Hierfür können den Blöcken im Simulink-Modell, unter dem Parameter „UserData“ zusätzliche Informationen hinterlegt werden. Diese Informationen werden bei der Transformation in eine Graphenbeschreibung, durch das MAnTrAS-System, mit übernommen.
Es ist zum Beispiel möglich den einzelnen Blöcken Informationen über ihre Verarbeitungszeit auf einem speziellen Prozessor zu übergeben. Mit diesen Verarbeitungszeiten in einem Modell lassen sich anschließend WCET[5]-Analysen durchführen 0. Durch die Transformation des Modells in eine Graphenbeschreibung lassen sich diese WCET-Analysen jedoch einfach auf ein graphentheoretisches Problem reduzieren. Aus dieser WCET-Analyse wird im Graphen eine Suche nach dem längsten Pfad.
Da Zyklen bei der Suche, nach dem längsten Pfad, ein Problem darstellen, müssen diese im Vorfeld durch geeignete Transformationen entfernt werden.
Es lassen sich neben dem genannten Beispiel eine Vielzahl anderer Informationen den Simulink-Blöcken hinzufügen. Dadurch ist es möglich die unterschiedlichsten Analysen eines Systems in die modellbasierte Entwicklung zu übernehmen.

## 5.2 Transformationen

Analysen lassen die Struktur eines Graphen unberührt, eine Transformation hingegen verändert Blöcke, ihre Eigenschaften oder Verbindungen untereinander und wirkt sich dabei gege-benenfalls auf das Verhalten des Modells aus. Der folgende Abschnitt beschreibt einige mit MAnTrAS umgesetzte Transformationen.

[5] Worst Case Execution Time: Die längste Zeit die ein Programmteil zur Ausführung benötigt.

### 5.2.1 Connection Normalization

Diese Transformation wird durchgeführt, um eine Vereinfachung des Simulink-Systems zu erreichen. Normalerweise sind im Simulink-Modell ausgehende Verbindungen als *1:n*-Beziehungen dargestellt. Das hat zur Folge, dass zwei miteinander verbundene Blöcke nicht über dieselbe Verbindung miteinander verbunden sind. Abbildung 4 stellt dieses Problem und die Auflösung schematisch dar.

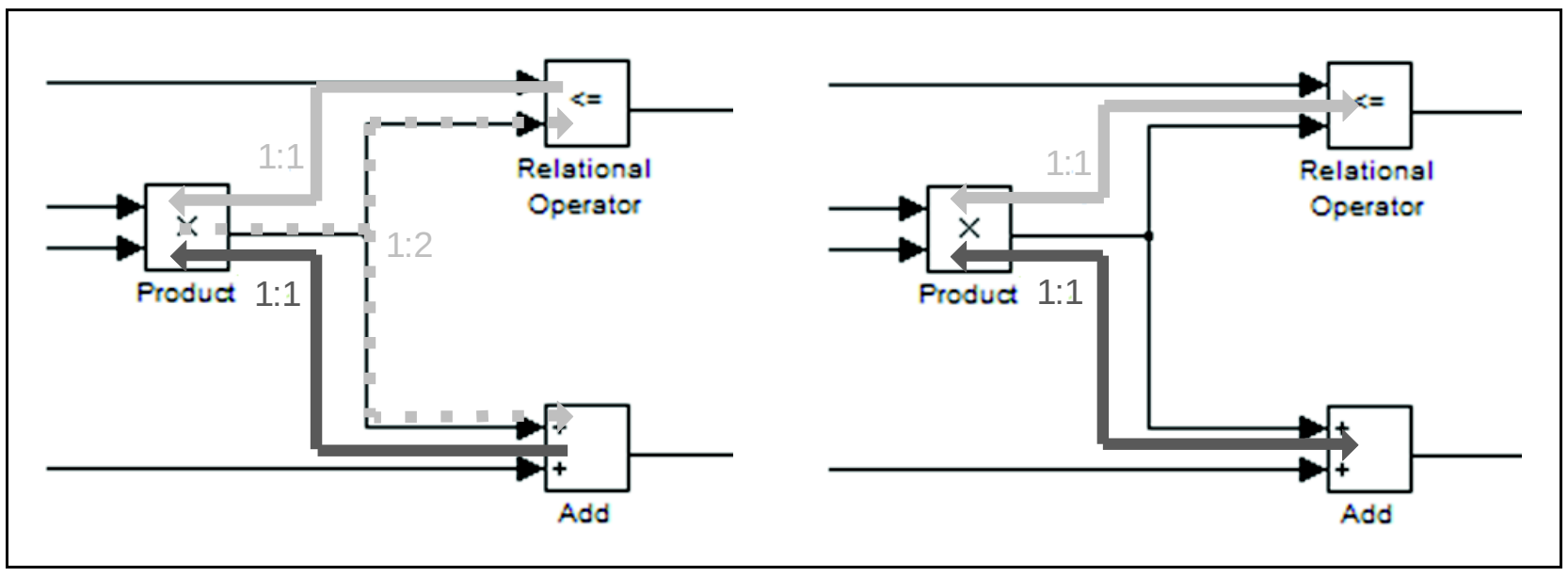

Abbildung 4: Das Prinzip der Normalisierung, die zwei unidirektionalen 1:1-Beziehungen (links dunkel- und hell-grau) und die unidirektionalen 1:2 Beziehung (links hell-grau gestrichelt) werden zu zwei bidirektionale 1:1-Beziehungen umgewandelt (recht dunkel und hell-grau).

Die hier durchgeführte Normalisierung verwendet wieder das "Visitor-Pattern", wobei dieses Mal alle Blöcke und Subsysteme eines Modells verarbeitet werden. Der Algorithmus durchsucht die Blöcke nach 1:*n*-Beziehungen und entfernt sie. Anschließend werden die Verbindungen zwischen den Blöcken in Form von *1:1*-Beziehungen neu gebildet.
Diese Normalisierung bringt der Vorteil, dass zwei Blöcke immer über eine einzige Verbindung miteinander verbunden sind, was die Erstellung von Algorithmen vereinfacht.

### 5.2.2 Abflachung der Hierarchie

Bei der Analyse sehr komplexer Systeme mit stark hierarchischer Struktur kann eine Auflösung der Hierarchien sinnvoll sein. Einerseits wird die Umsetzung von Algorithmen vereinfacht, da bei ihrer Erstellung Hierarchien nicht mehr berücksichtigt werden müssen. Andererseits können unter gewissen Umständen Algorithmen schneller abgearbeitet werden, womit deren Laufzeit verringert wird.
Die Abflachung der Hierarchien ist mithilfe eines "Visitor-Pattern" umgesetzt. Der Besucher arbeitet dabei alle Ports (Inports und Outports) eines Subsystems ab. So werden für Eingangsports (Ausgangsports) sowohl die internen Nachfolger (Vorgänger), als auch die externen Vorgänger (Nachfolger) ermittelt und die Vorgänger- und Nachfolgerlisten dieser Blöcke angepasst. Anschließend werden die Ports entfernt. Die im System verbleibenden Blöcke werden schließlich auf eine Hierarchieebene übertragen.

### 5.2.3 Substitution

Bei einer Substitution werden Blöcke oder Blockgruppen mit bestimmten Eigenschaften ersetzt oder modifiziert. Diese Transformation von einzelnen Blöcken lässt sich sehr einfach mit der vorgestellten API umsetzen: hier bietet sich wiederum das "Visitor-Pattern" an. Dieser iteriert über alle Blöcke eines bestimmten Typs und untersucht diesen auf Übereinstimmung mit dem gesuchten Block. Ist ein entsprechender Block gefunden wird dieser ersetzt oder eine spezifische Eigenschaft modifiziert.
Abbildung 5 zeigt Beispielhaft eine einfache Substitution, bei der „Gain"-Blöcke durch eine Multiplikation mit einer Konstante ersetzt werden. Diese Substitution hat keine Auswirkungen auf die Funktionalität eines Modells, zeigt aber, wie leicht mit MAnTrAS Modellrichtlinien umgesetzt werden können.

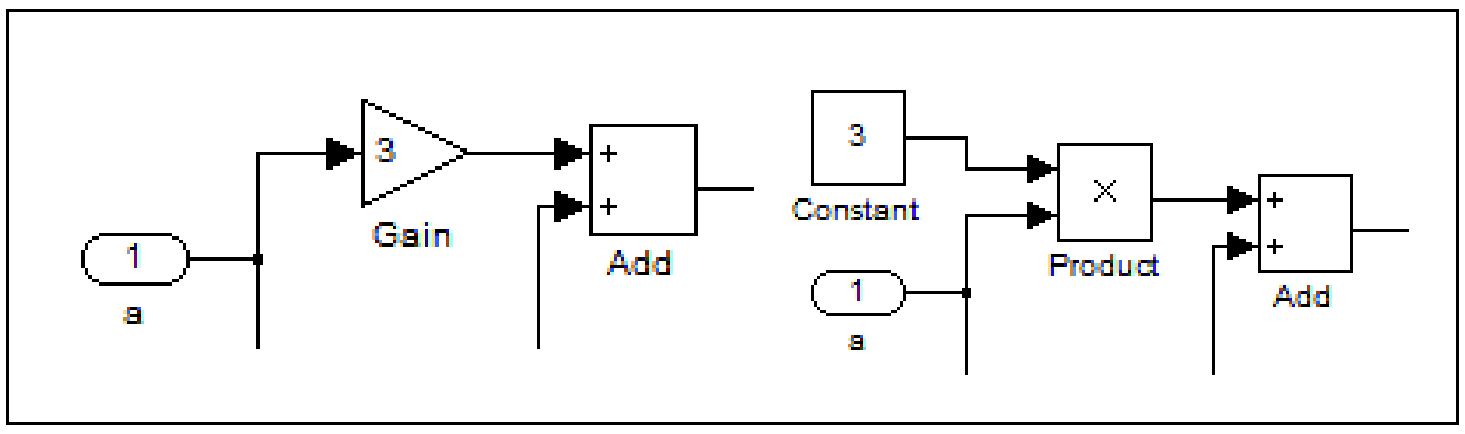

Abbildung 5: Einfache Substitution eines Blocks: der „Gain"-Block links wird durch eine Multiplikation (Produkt) mit einer Konstante (Constant) ersetz.

## 6 Fazit

In dieser Arbeit konnte aufgezeigt werden, dass es viele Einsatzbereiche gibt, in welchen Analysen und Transformationen von Modellen sehr nützlich sind. Hierfür bietet MAnTrAS eine API um solche Algorithmen, welche meist auf die Graphentheorie basieren, einfach zu implementieren. Die bisherigen Arbeiten [2-5] legen sehr starken Fokus auf die Qualitätsmessung durch erkennen von Mustern. Unser Ansatz geht einen entscheidenden Schritt weiter, da wir eine generische Programmierschnittstelle für MATLAB bereitstellen.
Durch die gezeigte Objektorientierung und das darauf aufbauende „Visitor-Pattern", anders als etwa in MatlabBGL [1], ist die Implementierung einfacher zu realisieren als über die native MATLAB/Simulink Schnittstelle. Durch dieses Framework lässt sich der Workflow für eine schnelle und einfach zu realisierende Modellanalyse und –transformation umsetzen. Durch die Automatisierung, welche das Framework bietet, wird das Risiko von Fehlern, im Gegensatz zu einer manuellen Bearbeitung, reduziert.
MAnTrAS ist dabei so flexibel, dass sich vielfältige Einsatzmöglichkeiten ergeben. Für zukünftige Anwendungen lassen sich beispielsweise eigene Algorithmen, zuvor definierte Transformations- als und Abfragesprachen implementieren.

## 7 Referenzen